\documentclass[aps,prl,twocolumn,superscriptaddress]{revtex4}
\usepackage{graphicx}
\usepackage{amsmath}
\usepackage{amssymb}
\usepackage{colordvi}
\usepackage{mathrsfs}
\usepackage{bm}
\usepackage{verbatim}
\usepackage{dcolumn}
\usepackage{epsfig}
\usepackage{subfigure}
\usepackage{array}
\usepackage{booktabs}
\usepackage{color}

\begin{document}

\title{Quantum Anomalous Hall Effect and Giant Rashba Spin-Orbit Splitting in Compensated \textit{n}--\textit{p}-Codoped Graphene}

\author{Xinzhou Deng}
\affiliation{ICQD, Hefei National Laboratory for Physical Sciences at Microscale, and Synergetic Innovation Center of Quantum Information and Quantum Physics, University of Science and Technology of China, Hefei, Anhui 230026, China.}
\affiliation{CAS Key Laboratory of Strongly-Coupled Quantum Matter Physics, and Department of Physics, University of Science and Technology of China, Hefei, Anhui 230026, China.}

\author{Hualing Yang}
\affiliation{School of Chemistry and Materials Science, Shanxi Normal University, Linfen, Shanxi 041004, China.}

\author{Shifei Qi}
\email[Correspondence author:~~]{qisf@sxnu.edu.cn}
\affiliation{School of Chemistry and Materials Science, Shanxi Normal University, Linfen, Shanxi 041004, China.}
\affiliation{ICQD, Hefei National Laboratory for Physical Sciences at Microscale, and Synergetic Innovation Center of Quantum Information and Quantum Physics, University of Science and Technology of China, Hefei, Anhui 230026, China.}

\author{Xiaohong Xu}
\affiliation{School of Chemistry and Materials Science, Shanxi Normal University, Linfen, Shanxi 041004, China.}

\author{Zhenhua Qiao}
\email[Correspondence author:~~]{qiao@ustc.edu.cn}
\affiliation{ICQD, Hefei National Laboratory for Physical Sciences at Microscale, and Synergetic Innovation Center of Quantum Information and Quantum Physics, University of Science and Technology of China, Hefei, Anhui 230026, China.}
\affiliation{CAS Key Laboratory of Strongly-Coupled Quantum Matter Physics, and Department of Physics, University of Science and Technology of China, Hefei, Anhui 230026, China.}

\begin{abstract}
  Quantum anomalous Hall effect (QAHE) is a fundamental quantum transport phenomenon in condensed matter physics. Until now, the only experimental realization of the QAHE has been observed for Cr/V-doped (Bi,Sb)$_2$Te$_3$ but at extremely low observational temperature, thereby limiting its potential application in dissipationless quantum electronics. Employing first-principles calculations, we study the electronic structures of graphene codoped with 5\textit{d} transition metal and boron (B) atoms based on a compensated \textit{n}--\textit{p} codoping scheme. Our findings are as follows. 1) The electrostatic attraction between the \textit{n}- and \textit{p}-type dopants effectively enhances the adsorption of metal adatoms and suppresses their undesirable clustering. 2) Hf--B and Os--B codoped graphene systems can establish long-range ferromagnetic order and open nontrivial band gaps because of the spin-orbit coupling with the non-vanishing Berry curvatures to host the QAHE. 3) The calculated Rashba splitting energy in Re--B and Pt--B codoped graphene systems can reach up to 158 and 85~meV, respectively, which is several orders of magnitude higher than the reported intrinsic spin-orbit coupling strength.
\end{abstract}

\maketitle

\textit{Introduction---.} Significant discoveries related to the quantum anomalous Hall effect (QAHE), which is a new quantum state of matter in condensed matter physics, have been reported in the recent years. The QAHE exhibits quantized Hall conductance in the absence of an external field, which arises from strong spin-orbit coupling combined with breaking of the time-reversal symmetry due to intrinsic magnetization~\cite{1-HaldanePRL1988,2-Weng2015,3-Ren-review2016,4-LiuAnnuRevCondMattPhys2016}. It is a topologically nontrivial phase characterized by a finite Chern number and chiral edge states within the bulk band gap. The chiral edge states are robust against backscattering and are promising for application to devices that require low-power consumption.

To date, the QAHE has been theorized to occur in systems such as Mn-doped HgTe quantum wells~\cite{5-Liu-Hg1-yMnyTePRL2008}, thin-film topological insulators (TIs)~\cite{6-Yu-Science2010}, silicene~\cite{7-Ezawa2012,8-ZhangPRB2013,9-ZhangSciRep2013}, half-hydrogenated Bi honeycomb monolayers~\cite{10-Liu-VQAHBi-PRB2015}, and graphene-based systems~\cite{11-Qiao-PRB2010,12-Qiao-PRB2012,13-Ding-PRB2011,14-Zhang-PRL2012,15-Qiao-PRL2014}. Graphene and TIs are candidate materials for engineering the QAHE because of their unique linear Dirac dispersion~\cite{16-Geim-NatMater2007,17-Zhang-Nature2005} and their mature technologies for growing samples. TIs are superior to graphene because they have stronger spin-orbit coupling (SOC), thereby narrowing the search for suitable materials for realizing the QAHE. An effective way to induce magnetism in TIs is to dope them with magnetic atoms~\cite{6-Yu-Science2010,18-Hor-PRB2010,19-Niu-APL2011,20-Haazen-APL2012,21-Jungwirth-RevModPhys2006}. The QAHE was first observed experimentally in Cr- and V-doped thin films at extremely low temperatures~\cite{22-Chang-Science2013,23-Checkelsky-NatPhys2014,24-Kou-PRL2014,25-Chang-NatMater2015}. Recently, the QAHE at temperatures above 50 K by using the \textit{n}--\textit{p} codoping in Sb$_2$Te$_3$ by means of vanadium--iodine has been theoretically proposed~\cite{26-Qi-PRL2016}, which has inspired the great experimental advance in exploiting hight-temperature QAHE~\cite{HeKe-AdvancedMaterials}.

Meanwhile, graphene has become a competitive candidate material for achieving QAHE because of its extraordinary electrical property, unique honeycomb lattice, and relatively mature technologies of sample growth~\cite{16-Geim-NatMater2007,17-Zhang-Nature2005}. However, pristine graphene possesses a very small nontrivial band gap because of its extremely weak intrinsic SOC~\cite{27-Yao-PRB2007,28-Gmitra-PRB2009}. Therefore, various measures such as hydrogen deposition~\cite{29-CastroNeto-PRL2009} have been proposed to enhance the extrinsic SOC in graphene. In particular, periodic adsorption of transition metal (TM) atoms has been suggested theoretically as an effective way to enhance SOC in graphene~\cite{11-Qiao-PRB2010,30-Hu-PRL2012}. However, periodic adsorption of TM atoms on graphene is currently very difficult to realize experimentally. A subsequent study found that random adsorption of TM atoms can eliminate the intervalley scattering associated with periodic adsorption~\cite{31-Jiang-PRL2012}. However, the adsorption energies of TM atoms on graphene are usually very low and the TM atoms tend to form clusters even at low temperatures~\cite{DopingCluster1,DopingCluster2}, indicating that the long-range ferromagnetic (FM) order may not survive as expected. Given the aforementioned problems, it is important to find new strategies to solve them. Recently, it was confirmed that \textit{n}--\textit{p} codoping is effective for enhancing the adsorption of TM atoms on graphene, leading to FM graphene~\cite{32-Qi-Carbon2013,33-Zhang-Carbon2015,34-Zhang-ApplSurfSci2014,xinzhouPRB2017}. Furthermore, the \textit{n}--\textit{p} codopants can help preserve the Dirac nature of the charge carriers.

In this article, by using first-principles calculations in the realm of density functional theory, we propose a versatile approach to achieving the QAHE based on a compensated \textit{n}--\textit{p} codoping scheme. This method was initially introduced to achieve \textit{p} doping in ZnO~\cite{35-Yamamoto-JpnJApplPhys1999,36-Wang-PRL2003} and to narrow the band gap of TiO$_2$~\cite{37-Gai-PRL2009,38-Zhu-PRL2009}, and more recently it was proposed in the context of diluted magnetic semiconductors~\cite{32-Qi-Carbon2013,33-Zhang-Carbon2015,xinzhouPRB2017,39-Xu-NewJPhys2006,40-Zhu-PRL2008}, which can also significantly enhance the adsorption of TM atoms on graphene. Meanwhile, the successful fabrication of B-substituted graphene also inspired our proposal~\cite{boronsubgraphene}. Therefore, we study the electronic and topological behaviors of 5\textit{d} TM--B codoped graphene systematically by utilizing \textit{n}--\textit{p} codoping. First, we find that the electrostatic attraction between the \textit{n}- and \textit{p}-type dopants effectively enhances the adsorption of the metal adatoms and suppresses their undesirable clustering. The results reveal that the magnetic coupling for all 5\textit{d} TM--B pairs shows a Ruderman-Kittel-Kasuya-Yosida type spatial fluctuation, with only Hf and Os being able to form long-range FM order. When SOC is considered, Hf--B and Os--B codoping are both able to open up a global band gap to facilitate the QAHE with the non-vanishing Berry curvatures. Next, we calculated the Rashba splitting energy in Re--B and Pt--B codoped graphene systems to be around 158 and 85~meV, respectively, which is several orders of magnitude larger than the reported intrinsic SOC strength.

\textit{Computational Methods---.} The calculations were performed using the projector augmented-wave formalism of density functional theory~\cite{41-blochl-PRB1994} as implemented in the Vienna ab initio simulation package (VASP)~\cite{42-Kresse-PRB1994,43-Kresse-PRB1993,44-Kresse-ComputMaterSci1996}. For exchange correlation, we used the Perdew-Burke-Ernzerhof~\cite{45-Perdew-PRL1996,46-Perdew-PRL1997} functional. The graphene sheet was modeled (i) with a 4$\times$4 supercell with a 20~{\AA} vacuum in the vertical direction for the adsorption calculations and (ii) with a 7$\times$7 supercell with a 20~{\AA} vacuum to estimate the magnetic interaction between two TM--B pairs. All our calculations used the optimized lattice constant of graphene with a value of $a_0=2.46$~{\AA}, which agrees with the experimental value. When optimizing the geometry, the positions of all atoms were allowed to relax and the atomic structures were optimized fully until the Hellmann-Feynman forces on each ion were less than 0.02~eV/{\AA}. The plane-wave energy cut-off was set as 500~eV with an energy precision of 10$^{-4}$~eV. For the 4$\times$4 (resp.\ 7$\times$7) supercell, the Brillouin zone was sampled using a 15$\times$15 (resp.\ 5$\times$5) $\Gamma$-centered k-point grid. The Gaussian smearing method was used with a smearing width of 0.1~eV. The adsorption energies of the 5\textit{d} TM adatoms on the B-substituted graphene sheet were estimated using
\begin{eqnarray}
E_{\rm{ad}} = E_{\rm{tot}} - E_{\rm{atom}} - E_{\rm{gra+B}}, \nonumber
\end{eqnarray}
where $E_{\rm{tot}}$, $E_{\rm{atom}}$, and $E_{\rm{gra+B}}$ are the energies of the 5\textit{d} TM--B codoped graphene system, an isolated TM atom, and B-substituted graphene, respectively.

\textit{Results and Discussion---.}
\begin{table}
	\centering
	\scriptsize
	\renewcommand{\arraystretch}{1.1}
	\begin{tabular}{>{\centering}p{.04\textwidth} >{\centering}p{.13\textwidth} >{\centering}p{.025\textwidth} >{\centering}p{.025\textwidth} >{\centering}p{.035\textwidth} >{\centering}p{.035\textwidth}>{\centering}p{.04\textwidth}}
		\toprule[1.2pt]
		&  \tabularnewline
		Atom        &  Site       &  \multicolumn{2}{c}{Move}  & & \multicolumn{2}{c}{$M (\rm{\mu_B})$ }  \tabularnewline
		&  \tabularnewline
		\midrule[.5pt]
		&  \tabularnewline
		            &  bridge          &no   &(yes) & & 0.56 &             \tabularnewline
		Hf          &  hollow          &no   &(no)  & & 2.71 &(3.11)       \tabularnewline
		            &  top             &no   &(no)  & & 0.37 &(1.50)       \tabularnewline
		            &  bridge          &no   &(yes) & & 2.00 &             \tabularnewline
		Ta          &  hollow          &no   &(no)  & & 3.82 &(3.68)       \tabularnewline
		            &  top             &yes  &(no)  & &      &(2.99)       \tabularnewline
		            &  bridge          &yes  &(yes) & &      &             \tabularnewline
		W           &  hollow          &no   &(no)  & & 3.29 &(2.31)       \tabularnewline
		            &  top             &yes  &(no)  & &      &(5.79)       \tabularnewline
		            &  bridge          &yes  &(yes) & &      &             \tabularnewline
		Re          &  hollow          &no   &(no)  & & 0.00 &(0.89)       \tabularnewline
		            &  top             &no   &(no)  & &      &             \tabularnewline
		            &  bridge          &yes  &(yes) & &      &             \tabularnewline
		Os          &  hollow          &no   &(no)  & & 1.00 &(1.97)       \tabularnewline
		            &  top             &no   &(yes) & & 3.00 &             \tabularnewline
		            &  bridge          &yes  &(no)  & &      &(0.95)       \tabularnewline
		Ir          &  hollow          &no   &(yes) & & 1.72 &             \tabularnewline
		            &  top             &no   &(yes) & & 1.93 &             \tabularnewline
		            &  bridge          &no   &(no)  & & 0.00 &(0.00)       \tabularnewline
		Pt          &  hollow          &yes  &(yes) & &      &             \tabularnewline
		            &  top             &yes  &(yes) & &      &             \tabularnewline
		            &  bridge          &yes  &(no)  & &      &(0.86)       \tabularnewline
		Au          &  hollow          &yes  &(no)  & &      &(0.84)       \tabularnewline
		            &  top             &no   &(no)  & & 0.00 &(0.85)       \tabularnewline
		&  \tabularnewline
		\bottomrule[1.2pt]
	\end{tabular}
	\caption[]
	{Magnetic moments ($M$) of TM atoms (TM = Hf, Ta, W, Re, Os, Ir, Pt, and Au) adsorbed on B-substituted (pristine) graphene at three possible adsorption sites. A blank entry indicates that the corresponding configuration is unstable.}\label{table1}
\end{table}
We begin by studying the adsorption of a single 5\textit{d} TM adatom on the graphene monolayer. We consider three highly symmetric adsorption sites, namely a carbon-atom top site, a carbon-carbon bridge site, and a hollow site. From Table~\ref{table1}, we find that two stable adsorption sites (data in parentheses in Table~\ref{table1}) are obtained for Hf, Ta, W, and Re, whereas only one adsorption site is obtained for Os, Ir, and Pt favor. In the case of an Au adatom on graphene, three stable adsorption sites (one bridge, one hollow, and one top) are obtained. Regarding the magnetic properties of a 5\textit{d} TM adatom on pristine graphene, no magnetic moment appears in the case of Pt-adsorbed graphene. However, because of the various adsorption configurations, various magnetic moments are obtained with the other 5\textit{d} TM adatoms adsorbed on graphene. These differing configurations and magnetic moments reflect the differing charge transfer when these 5\textit{d} TM adatoms are adsorbed onto the graphene at different sites. Our results also show that the hollow site is the most favored adsorption site except for Pt and Ir, for which the bridge site is preferred. This finding agrees with the findings of previous theoretical studies~\cite{14-Zhang-PRL2012}. Our calculations also show that the largest adsorption energy among the various 5\textit{d} TM adatoms (namely that for Ta) is less than 2~eV on pristine graphene. This relatively weak binding between graphene and each of these 5\textit{d} TM adatoms, as well as the small diffusion barrier, results in fast adatom migration and clustering. We show below that codoping with B is effective for suppressing these undesirable effects.

\begin{figure}
	\includegraphics
	[width=8.5cm,angle=0]{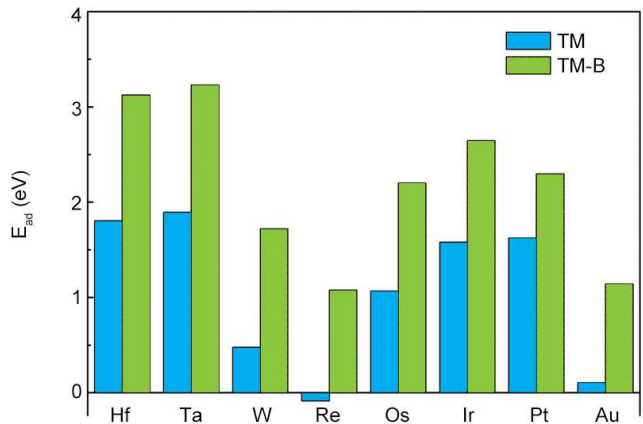}
	\caption{ (color online) Adsorption energies of 5\textit{d} TM adatoms on pristine graphene (blue) and B-codoped graphene (green).
	}
	\label{figEad}
\end{figure}

As mentioned above, the relatively low adsorption energies of TM-doped graphene cannot prevent the fast migration of adatoms and clustering with other adatoms adsorbed on graphene. The 5\textit{d} TM adatoms are \textit{n}-type dopants in graphene, whereas B adatoms (occupying the same row as carbon in the periodic table with just one less electron) are \textit{p}-type dopants. Therefore, a 5\textit{d} TM adatom will locate close to a B adatom because of the strong \textit{n}--\textit{p} electrostatic attraction between them. Previous studies have also shown that B codoping is effective for suppressing the present undesirable effects~\cite{32-Qi-Carbon2013,xinzhouPRB2017}. To show this is indeed the case, we calculated the adsorption energies of the 5\textit{d} TM adatoms from Hf to Au after B codoping. As shown in Fig.~\ref{figEad}, the adsorption energies all exhibit a $\sim$1~eV enhancement compared with those on pristine graphene, indicating a strong attractive interaction between TM and B. We found that most of the 5\textit{d} TM adatoms yield the strongest binding at the hollow site after codoping except for Pt and Au, for which the bridge and top sites are preferred, respectively. Such strong attraction helps pin the TM adatoms close to B, thereby preventing the adatoms from migrating and clustering on the graphene. Regarding the magnetic moment of graphene codoped with B and 5\textit{d} TM adatoms, Table~\ref{table1} indicates that no magnetic moment appears when using Re, Pt, or W. In fact, B codoping results in more charge transfer, thereby changing the magnetic moment from that in the case of using only 5\textit{d} TM adatoms. Therefore, the abovementioned findings confirm that \textit{n}--\textit{p} codoping distributes 5\textit{d} TM adatoms uniformly and stably on B-substituted graphene.

\begin{figure}
	\includegraphics
	[width=8.5cm,angle=0]{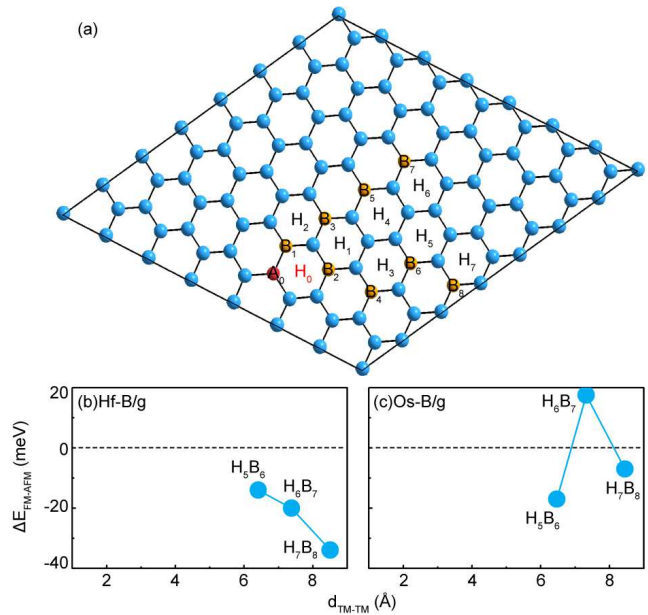}
	\caption{ (color online) (a) A 7$\times$7 graphene supercell. H$_0$-H$_7$ represent the hollow sites for TM adsorption, A$_0$ and B$_1$-B$_8$ represent the B-substituted ``A" site and ``B" sites of carbon atoms in graphene, respectively. (b)(c) Magnetic coupling between two TM--B pairs vs. their separation. One pair is fixed at H$_0$A$_0$ and the other one moves from H$_1$B$_2$ to H$_7$B$_8$, as illustrated in (a).
	}
	\label{figFMAFM}
\end{figure}

The results presented thus far show that 5\textit{d}-TM--B codoping enhances the stability of TM adatoms on graphene significantly, thereby providing the essential prerequisites to achieve enhanced SOC and to realize the QAHE in graphene. Ferromagnetism is another crucial requirement to realize the QAHE in \textit{n}--\textit{p} codoped graphene, which is why we next consider the magnetic interaction between two TM--B pairs in a 7$\times$7 graphene supercell. As shown in Fig.~\ref{figFMAFM}(a), one TM adatom is fixed at H$_0$ and paired with a B atom at A$_0$. The other TM adatom moves from H$_1$ to H$_7$, whereas its partner B codopant remained at its nearest neighbor. The magnetic interaction between the two TM--B pairs at a given separation is evaluated by calculating the total energy difference between the FM and the antiferromagnetic (AFM) configurations of the two TM moments. We find that the first to fourth nearest-neighbor configurations are unstable because of the surprisingly large attraction between the two 5\textit{d} TM adatoms even though their corresponding adsorption energies increase after B codoping. However, it remains possible to realize long-range ferromagnetism because configurations appear with larger separations between the two codopants. As shown in Fig.~\ref{figFMAFM}(b) and (c), we also find that both Hf--B and Os--B codoped graphene systems display FM coupling at sufficiently large TM-TM distance, indicating long-range FM order in these two systems. However, our calculations show no long-range FM order in the other types of 5\textit{d}-TM--B codoped graphene. Therefore, the Hf--B and Os--B codoped graphene systems are the most promising candidates for achieving the QAHE.

\begin{figure}
	\includegraphics
	[width=8.5cm,angle=0]{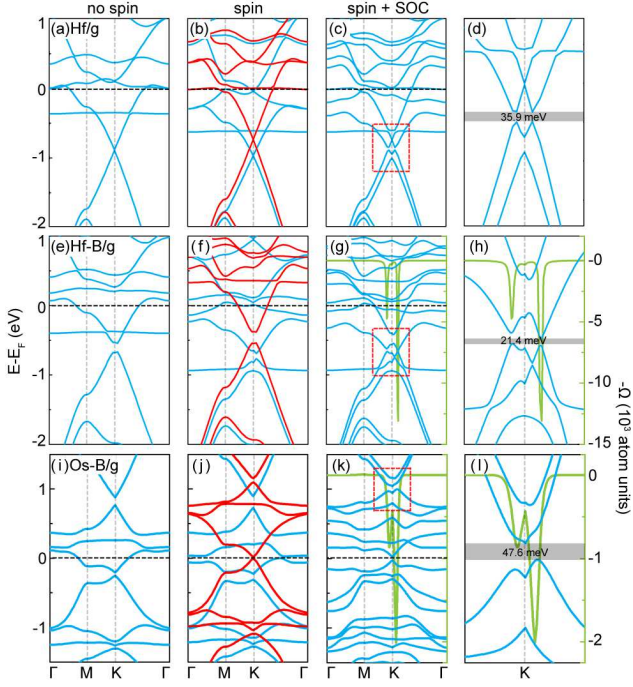}
	\caption{ (color online) Band structures of Hf, Hf--B, and Os--B/graphene systems: (a)(e)(i) no spin; (b)(f)(j) without SOC; (c)(g)(k) with SOC. The red and blue lines denote the spin-up and spin-down bands, respectively. The green lines represent the Berry curvature. (d)(h)(l): magnified views of the red rectangle in (c), (g), and (k), respectively.
	}
	\label{figHfHfOs}
\end{figure}

The 5\textit{d}-TM--B codoping indeed enhances the stability of TM adatoms significantly, and the Hf--B and Os--B codoped graphene systems can realize long-range ferromagnetism. Next, to see whether the QAHE can be realized in \textit{n}--\textit{p} codoped graphene, we calculate the band structures of Hf--B and Os--B codoped graphene. First, we examine Hf on graphene in a 4$\times$4 supercell with and without SOC. In Fig.~\ref{figHfHfOs}, panels (a) and (b) show the band structures of Hf-doped graphene without SOC, wherein the unique linear dispersion of pure graphene is preserved. In Fig.~\ref{figHfHfOs}(c), a gap of $\sim 35.9$~meV opens at the Dirac point when SOC is activated, indicating that this gap definitely originates from the SOC related to Hf adatom adsorption. The abovementioned results agree completely with those of Zhang et al.~\cite{14-Zhang-PRL2012}, indicating that our computational method is reliable. Next, we show in Figs.~\ref{figHfHfOs}(f) and (j) the band structures of Hf--B and Os--B coped graphene, respectively, without SOC. When magnetization is included, the spin-up (resp.\ spin-down) bands are shifted upward (resp.\ downward). When SOC is also activated, band gaps open at the crossing points between the spin-up and spin-down bands near the K point, as shown in Figs.~\ref{figHfHfOs}(g) and (k), with magnitudes of 21.4 and 47.6~meV, respectively. Therefore, the QAHE might be realized in Hf--B and Os--B codoped graphene.

\begin{figure}
	\includegraphics
	[width=8.5cm,angle=0]{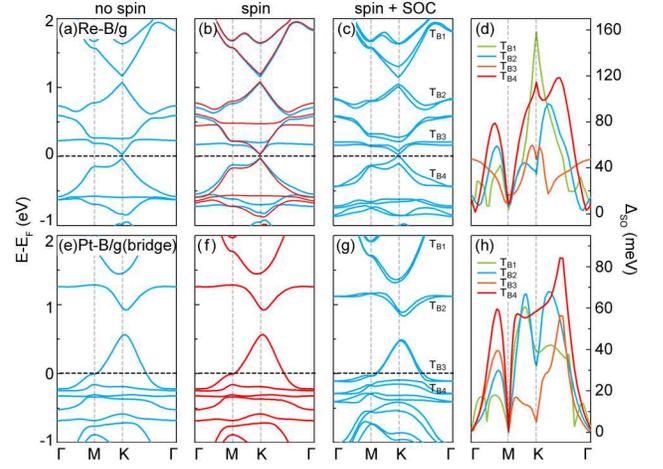}
	\caption{ (color online) Band structures of Re--B and Pt--B/graphene systems: (a)(e) no spin; (b)(f) without SOC; (c)(g) with SOC. (d)(h) Absolute values of spin-orbit (SO) splitting ($\Delta_{\rm{SO}}$) of T$_{\rm{B1}}$-T$_{\rm{B4}}$ in (c) and (g), respectively.
	}
	\label{figRePt-B}
\end{figure}

So far, we have shown that joint effect between SOC and ferromagnetic exchange field can open up band gaps in Hf--B and Os--B codoped graphene. Next, we investigate whether such band gaps can host the QAHE via Berry curvature calculations using the expression~\cite{Berry1,Berry2}
\begin{eqnarray}
\Omega(\bm{k})=-\sum_{n}f_n\sum_{n'\neq n} \dfrac{2 {\rm{Im}} \langle{\psi_{n{k}}|v_x|\psi_{n'{k}}\rangle} \langle{\psi_{n'{k}}|v_y|\psi_{n{k}}\rangle}}{(E_{n'}-E_n)^2}, \nonumber
\end{eqnarray}
where $n$, $E_n$, and $\psi_{nk}$ are the band index, eigenvalue, and eigenstate, respectively, of the $n$-th band. $v_{x,y}$=$\partial E/\partial k_{x,y}$ are velocity operators in the $x$ and $y$ directions within the film plane, and $f_n = 1$ for all $n$ bands below the band gap. Figures~\ref{figHfHfOs}(g) and (k) show the distributions of Berry curvature along the highly symmetric lines. Large negative peaks appear near the K points and vanish elsewhere, demonstrating nonzero Hall conductance. Therefore, our calculations confirm that these gaps can host the QAHE.

We also present the band structures of the Re--B and Pt--B codoped graphene systems even though they have no magnetism. The overall band structures of Re--B and Pt--B codoped graphene with and without SOC are shown in Fig.~\ref{figRePt-B}. With SOC, there is clearly no topologically nontrivial band gap. However, Figs.~\ref{figRePt-B}(c) and (g) show the existence of spin-orbit splitting. In contrast to Figs.~\ref{figRePt-B}(b) and (f) without SOC, the SOC interaction breaks the spin degeneracy of the bands of Re--B and Pt--B codoped graphene. In Figs.~\ref{figRePt-B}(d) and (h), we show the spin splitting $\Delta_{\rm{SO}}$ for the bands denoted as in Figs.~\ref{figRePt-B}(c) and (g) along the highly symmetric line $\Gamma$-M-K-$\Gamma$. These bands are directly relevant to the transport properties of \textit{n}--\textit{p} codoped graphene because their band energies are close to the Fermi level. The maximum spin splitting energy is roughly about 158 and 85~meV in the Re--B and Pt--B codoped graphene, respectively, which is much larger than the typical value for the Rashba spin splitting energy in conventional III--V and II--VI semiconductor quantum wells ($\textless$30~meV)~\cite{43-Kresse-PRB1993,44-Kresse-ComputMaterSci1996} and is fully comparable to the enhanced surface Rashba spin splitting energy ($\sim 100$~meV) in different graphene/substrate systems~\cite{19-Niu-APL2011,20-Haazen-APL2012,23-Checkelsky-NatPhys2014}. As a result, we reason that Re--B and Pt--B codoped graphene would be useful for designing new electronic materials. As for Ta--B, W--B, I--B, and Au--B codoped graphene, their band structures contain only trivial gaps or lack long-range FM order, indicating that the QAHE cannot be achieved.

\textit{Conclusions---.} In summary, through calculations based on density functional theory, we systematically studied the electronic and spintronic properties of 5\textit{d} TM--B codoped graphene based on a compensated \textit{n}--\textit{p} codoping scheme. The electrostatic attraction between the \textit{n}- and \textit{p}-type dopants effectively enhanced the adsorption of the metal adatoms and suppressed their undesirable clustering. The calculated Rashba splitting energy in the Re--B and Pt--B codoped graphene reached 158 and 85~meV, respectively, which is several orders of magnitude larger than the reported intrinsic SOC strength. Moreover, Hf--B and Os--B codoped graphene can support long-range FM and host the QAHE with non-vanishing Berry curvatures.

\textit{Acknowledgments---.} This work was financially supported by the National Key Research and Development Program (grant no.\ 2017YFB0405703), the National Natural Science Foundation of China (grant nos.\ 11104173, 61434002, and 51025101) and Sanjin Scholar of Shanxi. XD and ZQ also acknowledge the support of the China Government Youth 1000-Plan Talent Program and the National Key Research and Development Program (grant no.\ 2016YFA0301700). We are grateful to the supercomputing service of AM-HPC and the Supercomputing Center of USTC for providing the high-performance computing resources used in this study.

\end{document}